\documentstyle[twoside,fleqn,espcrc2]{article}

\def\be{\begin{equation}}
\def\ee{\end{equation}}
\def\ba{\begin{array}}
\def\ea{\end{array}}
\def\l{\label}
\def\refe#1{(\ref{#1})}
\def\T{{\rm \scriptscriptstyle T}}
\def\D{{\rm \scriptscriptstyle D}}

\newcommand{\AmS}{{\protect\the\textfont2
  A\kern-.1667em\lower.5ex\hbox{M}\kern-.125emS}}

\title{R-parity breaking and the grand unification program}

\author{Francesco Vissani\address{International Centre for Theoretical
        Physics, ICTP, 
        Strada Costiera 11, Trieste, Italy}\address{Istituto
        Nazionale di Fisica Nucleare, INFN, Sezione di Trieste} 
}
       
\begin{document}

\begin{abstract}
We present the study of the possibility to have 
R-parity breaking interactions in minimal SU(5).
An interesting scenario emerges in which the
R-parity breaking coupling $\lambda'_{333}$ is large, 
and its size is related to the value of the 
tau neutrino mass, assuming that supersymmetry 
is broken according to low energy supergravity framework. 
This scenario may also have implications 
on the issue of $b-\tau$ unification.
\end{abstract}

\maketitle

\section{Spectrum of minimal SU(5)}

The three gauge couplings constants of the Standard Model (SM), 
extrapolated assuming the existence 
of supersymmetric partners around the SU(2)$\times$U(1)
breaking scale, are consistent with the hypothesis
of SU(5) grand unification taking place
at energies $E={\cal O}(10^{16})$ GeV.

The matter and the Higgs fields can be arranged in supermultiplets,
with the usual gauge quantum number: $\bar 5_i$ and $10_i$ 
for the matter, where $i=1,2,3$ refers to the family replica;
$5_H$ and $\bar 5_H$ for the Higgs field giving mass to the ordinary
fermions;  $24_H$ for the Higgs field that breaks SU(5)
(the Higgs sector being the less certain from both the experimental
and theoretical point of view).
 
The Higgs 5-plets, beside  
the usual Higgs doublets, contain triplets
that are not part of the SM spectrum, 
whose exchange can mediate proton decay with strength
of the order of the Yukawa couplings.
The usual solution of this phenomenological impasse
is to assume that the triplets have a mass of the order
of SU(5) breaking scale, whereas the doublets have mass
of the order of SU(2) breaking scale (doublet-triplet splitting).

\section{R-parity violating couplings}

Since the Higgs fields $\bar 5_H$ is in the same gauge 
group representation
of the matter fields $\bar 5_i,$ beside the mass terms
$\bar 5_H \bar 5_i 10_j,$ and beside 
$\bar 5_H 5_H + \bar 5_H 24_H 5_H,$ 
giving rise to the $\mu$-term, we can write
the gauge group invariants [1-3,6-8] 
that break explicitly R-parity (or equivalently matter parity):
\be
\ba{l}
\Lambda_{ijk} \bar 5_i \bar 5_j 10_k \\
+ \bar 5_i (M_i + h_i 24_H ) 5_H .
\ea
\l{su5-superpot}
\ee
The terms in the two lines 
are quite different in character. 
We will refer to them as
``triliner matter couplings'' and ``matter-higgs mixings'' respectively.
Henceforth we will think of the bracketed term as a single effective term,
since, regarding phenomenological implications,
the role of the $24_H$ is just to communicate the SU(5) breaking. 

\subsection{Trilinear matter couplings}

If the first term in \refe{su5-superpot} is dominating,
we deduce that lepton- and baryon-number violating 
interactions are induced with the same strength.
Supersymmetric partners of quark fields can in this case mediate
proton decay, and therefore the size of the couplings $\Lambda_{ijk}$ 
is strongly limited in size from the available informations 
on proton stability: $\Lambda_{ijk} < {\cal O}(10^{-9}).$
Such a small couplings are not interesting for 
experimental search at accelerators.

The key hypothesis that forces this conclusion is the quark-lepton
symmetry, which implies that baryon- {\em and} lepton-violations 
appear at the same time.
Therefore, a conjectural observation of 
large R-parity breaking couplings 
should be related to asymmetries 
between quarks and leptons.
In the next section we illustrate how such an asymmetry 
can arise in the simple model at hand.

\subsection{Matter-higgs mixings}

The asymmetry between quarks and leptons can be 
related to the doublet-triplet splitting.
In fact, after the SU(5) breaking has taken place, 
the second term in eq.\ \refe{su5-superpot}
can be written together with the R-parity 
conserving analogue as follows:
\be
\ba{l}
(B^c, L_3)
\left( \ba{cc} m_\T& 0 \\ 0 & m_\D \ea \right) 
\left( \ba{c} {\cal T} \\ H_2 \ea \right) \\
+ ({\cal T}^c, H_1)
\left( \ba{cc} M_\T& 0 \\ 0 & M_\D \ea \right) 
\left( \ba{c} {\cal T} \\ H_2 \ea \right) ,
\ea
\ee
where we retained the mixing terms of the higgs with 
third generation matter fields, and introduced  four
effective masses for triplet and
doublet subspaces (we assume that, consistently with previous 
discussion, the trilinear matter couplings are negligible). 
It is possible to redefine
as superfield triplet ${\cal T}^c$ and as doublet $H_1$
the combinations:
\be
\ba{l}
m_\T B^c + M_\T {\cal T}^c \to \mu_\T {\cal T}^c \\
m_\D L_3 + M_\D H_1 \to \mu_\D H_1 ,
\ea
\ee
but this redefinition (called ``matter-higgs rotation'') 
will modify other terms in the lagrangian.
In particular, an R-parity 
violating coupling of the
$L_3 Q_3 B^c$ monomial will be generated out of the
bottom Yukawa term $H_1 Q_3 B^c$ with strength:
\be
\lambda_{333}'\sim y_b\ \theta_\D ,
\l{lambda333}
\ee
where $\theta_\D$ is the matter-higgs mixing angle and
$y_b$ the bottom Yukawa coupling.
Under the assumption that $M_\T$ is of the order
of the SU(5) breaking scale and the other masses are 
of the order of the electroweak scale
(a generalization of the doublet-triplet 
splitting hypothesis), 
the baryon-violating couplings are small and 
do not conflict with proton stability.
At the same time we can have values of $\lambda_{333}'$ coupling 
interesting for search at accelerators.

One phenomenological and one theoretical remarks are in order.
Let us notice that hadronic jets 
originating from a vertex $\lambda'_{333}$
are enriched in bottom quarks; furthermore, if the
angle $\theta_\D$ is small enough, the decay of the 
lighter supersymmetric particles will be slow and originate 
secondary vertices.
The other remark is that the bottom mass
is {\em smaller} than the tau Yukawa mass
at the unification scale after the matter-higgs rotation. 
This can be rephrased
by saying that the matter-higgs rotation breaks the 
SU(4) residual symmetry associated to the 
5-plet breaking pattern. 
A second possible effect on $b-\tau$ unification 
is discussed below.

The matter-higgs rotation has in general 
important effects also on 
the scalar sector of the theory [1,3-8].
To discuss this point properly one must assume 
a framework for supersymmetry breaking.
We consider here the low-energy supergravity
model, that, implying universality, guarantees the
absence of large flavour-changing neutral currents.
Let us write the part of the scalar potential that 
contains the Higgs doublet $h_1$ and the 
slepton doublet $\tilde l_3$ in the following manner:
\be
\ba{l}
V \ni (m_{L_3}^2+\delta m^2)\ |h_1|^2 + 
m_{L_3}^2\ |\tilde l_3|^2\ - \\[1ex] \nonumber
[B\cdot M_\D\ h_1 h_2 + (B+\delta B)\cdot m_\D\ 
\tilde l_3 h_2 + {\rm h.c.}].
\l{scalar-potential}
\ea
\ee
At the grand unification (or Planck) scale we have:
\be
\ba{c}
\delta B(M_X)=0 \\
\delta m^2(M_X)=0 ,
\ea
\ee
as a consequence of supergravity. 
Were this be true at all the scales,
the matter-higgs rotation would be harmless in the scalar sector. 
However, non-zero values of $\delta B$ 
and $\delta m^2$  will be induced by ${\cal O}(y_b^2)$ 
renormalization group effects [1,6-8]. 
[It is convenient to study the running before
the matter-higgs rotation is performed, since in this
basis the massive terms that break R-parity do not 
modify the evolution of the other terms.]
For this reason the potential in the rotated basis will contain 
{after SU(2)$\times$U(1) breaking the terms:
\be
V_{{\rm \scriptscriptstyle L \!\!\!\!\!\;/}}  \sim  \theta_\D \times 
\left[ 
\delta m^2\ h_1^* + \delta B\cdot \mu\ h_2 
\right] \tilde l_3 + 
{\rm h.c.} ,
\l{lepton-violating-part}
\ee 
implying a breaking of the tau lepton number. 
The sneutrino vacuum expectation value can be estimated as:
\be
{\langle \tilde \nu_3 \rangle}\sim v\ \theta_\D\ y_b^2 ,
\l{vev-estimate}
\ee
barring the possibility of cancellations 
among various contributions. 
As a consequence of  the sneutrino VEV,
the tau neutrino mixes with the zino, 
and become massive.
Equations \refe{lambda333} and \refe{vev-estimate}
imply a correlation between the R-parity
breaking couplings and the tau neutrino mass:
\be
\ba{l}
\lambda_{333}'\sim 0.001\times \\
\left[\frac{\theta_\D}{0.1\ {\rm rad.}} \right]^{1/2}
\!\!\!\times \left[\frac{m_{\nu_\tau}}{30\ {\rm eV}} \right]^{1/4}
\!\!\!\times \left[\frac{M_{\tilde Z}}{1\ {\rm TeV}} \right]^{1/4}.
\ea
\l{correlation}
\ee
that, with a tau neutrino in an interesting range for cosmological 
considerations, does not preclude the discovery at accelerators 
of R-parity breaking effects. [The domain of validity 
of formula \refe{correlation} is limited by the perturbativity 
of the Yukawa couplings, its accuracy depending mainly 
on the measured value of the bottom mass and on the 
estimation \refe{vev-estimate}.]

Finally we comment on the scenario in which 
the tau neutrino mass is large, ${\cal O}$(MeV). 
This possibility is related to the fact that the 
parameter $\theta_\D$ is arbitrary, and can thus be chosen to 
saturate the experimental bound.
Among the consequences of this assumption
we remark the following:
$(i)$ Of course $m_{\nu_\tau}$ may be measured in $\tau$ decay experiments.
$(ii)$ Such a heavy neutrino 
should be unstable to avoid cosmological bounds.
[A proper discussion of this issue would requires to quantify 
the mixing with the other neutrinos, and will not be 
addressed here.]
$(iii)$ Regarding search at accelerators, 
it is important to remark 
that the correlation \refe{correlation} 
does not affect dramatically the size of the R-parity breaking 
coupling. 
$(iv)$ In the optimistic case in which the
effects of $\lambda_{333}'$ would show up at 
accelerators, the model described above for the neutrino mass 
should be considered as an alternative to the see-saw model.
$(v)$ The vacuum expectation value \refe{vev-estimate},
combining with the R-parity breaking coupling \refe{lambda333},   
will give a contribution to the bottom mass
of the order $\delta m_b \sim m_b\ (\theta_\D\ y_b)^2$
(its actual size and sign depending on the values of the 
supersymmetric parameters), that may be relevant for
$b-\tau$ unification.
This is larger than the ${\cal O}(\theta_\D^2)$ 
contribution from matter-higgs rotation if 
tan$\beta=\langle h_2^0\rangle/\langle h_1^0\rangle$  is large.
\vfill
I thank Alexei Yu.\ Smirnov for the pleasant collaboration
upon which this work is based, 
the Organizers of SUSY96 for invitation and partial support,
and Goran Senjanovi\'c for 
beautiful scientific discussions and for discovering
{\em Planet X.}  

\end{document}